\documentclass[twocolumn,aps,prl,superscriptaddress,footinbib,amsmath,amssymb,floatfix,10pt]{revtex4-2}

\usepackage{physics,bm,graphicx,lipsum,hyperref}
\newcommand{\br}{{{\bf{r}}}}
\usepackage[toc,page,titletoc]{appendix}

\newif\ifsm
\smtrue

\begin{document} 
\title{Dimer-projection contact and the clock shift of a unitary Fermi gas}

\newcommand{\Toronto}{Department of Physics and CQIQC, University of Toronto, Ontario M5S~1A7, Canada}
\newcommand{\SYS}{School of Physics and Astronomy, Sun Yat-Sen University (Zhuhai Campus), Zhuhai 519082, China}
\newcommand{\JQI}{Joint Quantum Institute, NIST / University of Maryland, College Park, Maryland 20742, USA}
\newcommand{\Heidelberg}{Institut f\"ur Theoretische Physik, Heidelberg University, 69120 Heidelberg, Germany}

\author{Kevin G. S. Xie}
\thanks{These two authors contributed equally.}
\author{Colin J. Dale}
\thanks{These two authors contributed equally.}
\author{Kiera Pond Grehan}
\author{Maggie Fen Wang}
\affiliation{\Toronto}
\author{Tilman Enss}
\affiliation{\Heidelberg}
\author{Paul S. Julienne}
\affiliation{\JQI}
\author{Zhenhua Yu}
\affiliation{\SYS}
\author{Joseph H. Thywissen}
\affiliation{\Toronto}

\begin{abstract}
Understanding the dynamics of short-range correlations is a central challenge in strongly interacting Fermi gases. In ultracold gases, these correlations are quantified by the contact parameter, yet measurements to date have been limited to equilibrium systems or relatively slow, global dynamics. 
Here, we introduce a rapid spectroscopic technique based on projection of the interacting state onto an alternate scattering channel with a low-lying dimer state. We demonstrate contact measurements on the microsecond timescale -- faster than the inverse Fermi energy. 
Using $^{40}$K near a broad $s$-wave Feshbach resonance, we show that the strength of the dimer-projection feature scales proportionally with the contact parameter extracted from the high-frequency tail of radio-frequency spectroscopy, in agreement with coupled-channels calculations. 
Analysis of the spectra further reveals that the dimer feature provides the dominant contribution to the clock shift of the unitary Fermi gas, allowing the first experimental bound on this quantity. 
The observed deviations from universal predictions highlight the importance of multichannel effects. 
Our results open new avenues for studying contact correlators, hydrodynamic attractors, and quantum critical behavior. 
\end{abstract}
\maketitle

{\em Introduction.}
Pair correlations play a central role in understanding strongly interacting quantum gases. In ultracold systems with large scattering lengths, the contact parameter $C$ quantifies the strength of short-range pair correlations. Contact relations directly link $C$ to a variety of physical properties, including thermodynamics, the high-momentum tail, spectral features, and the breaking of scale invariance \cite{Tan:2008a,Tan:2008b,Tan:2008c,Braaten:2008ez,Zhang:2009kq,Braaten:2012gh,Werner:2012um}.
While these properties are well established in equilibrium, experiments have yet to realize proposals that connect the time-dependence of $C$ to equilibration of Fermi gases \cite{fujii2018hydrodynamics, nishida2019, enss2019bulk, hofmann2020, fujii2024, Mazeliauskas2025} or quench dynamics of Fermi \cite{Son:2010bo,Cazalilla:2019} and Bose gases \cite{Natu:2013bi,Bohn:2015,PhysRevA.99.023623}. 
To probe local relaxation processes using contact correlators, the measurement time must be shorter than the characteristic collisional timescale, which for a unitary gas is on the order of the Fermi time, $\tau_F = \hbar/E_F$, where $E_F$ is the Fermi energy \cite{Makotyn2014,Cetina:2016,Eigen:2018,enss2019bulk,Cazalilla:2019,Vale:2021,Navon:2025a,Haibin:2025}. For typical ultracold-gas experiments, $\tau_F$ is on the order of microseconds. 

The contact parameter can be measured via radiofrequency (rf) spectroscopy 
\cite{SpectrosocpyReview2021}. A typical protocol begins with a mixture of two hyperfine states, $\ket{1}$ and $\ket{2}$, and measures the rf-induced transfer rate $\Gamma$ from $\ket{2}$ to $\ket{3}$, a weakly interacting state that is initially unpopulated. Two key aspects of $\Gamma$ have been linked theoretically to $C$: the high-frequency tail (HFT) \cite{Haussmann:2009,Pieri:2009jp,Braaten2010,Randeria2010} and the $\Gamma$-weighted average frequency shift $\Delta$ \cite{Baym:2007ki,Punk2007,Zhang:2008dc,Zhang:2009kq,Braaten2010,Langmack:2012hr,Fletcher:2017,Beugnon:2021}, commonly referred to as the ``clock shift''. While both observables have been treated generally, here we will consider the case of a spin-balanced ($N_1 = N_2 = N/2$) Fermi gas at unitarity ($a_{12}^{-1}=0$). 
The HFT is the form taken by $\Gamma$ for $\omega \gg E_F/\hbar$, which in the zero-range case is \cite{Braaten2010}
\begin{equation} \label{eq:GammaAsymptotic} \Gamma_\mathrm{HFT}^\mathrm{zr}(\omega) \approx \frac{\hbar^{1/2} \Omega_{23}^2}{8 \pi m^{1/2}} \frac{\omega^{-3/2}}{1 + \omega/\omega_a} C\,, \end{equation} 
where $\Omega_{23}$ is the $\ket{2} \to \ket{3}$ Rabi frequency, $\omega$ is the rf detuning from the single-particle $2 \to 3$ resonance, and $\omega_a \equiv \hbar/m a_{13}^2$ is the frequency scale at which final-state effects become apparent. The signature $C \omega^{-3/2}$ feature has been used widely to measure the contact parameter \cite{Stewart:2010fy,Sagi2012,Bardon:2014,Sagi:2018} and the $\omega > \omega_a$ roll-over has been observed \cite{mukherjee2019}. 

\begin{figure}[tb!]
\centering
\includegraphics[width=\columnwidth]{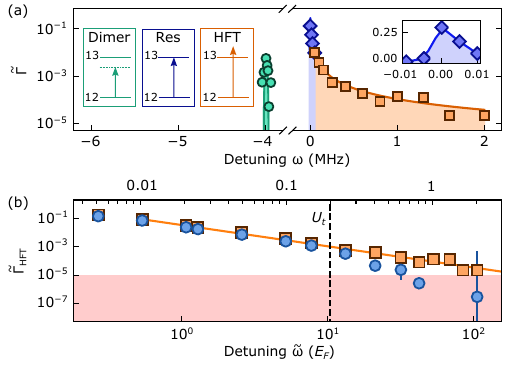}
\caption{{\em Spectral regimes.} {(a)} The dimensionless rf transfer rate $\widetilde{\Gamma} = E_F \alpha /(\pi \hbar \Omega_{23}^2 t_\mathrm{rf})$ is shown versus rf detuning $\omega$ from the single-particle resonance in three regimes: dimer feature (green circles), near-resonant transfer (blue diamonds) and HFT transfer (orange squares). Lines are intended to be guides for the eye. {(b)} HFT transfer is shown versus frequency with a $\omega^{-3/2}$ fit as the solid orange line. Here, $U_t$ (dashed line) denotes the trap depth. Blue circles use a positive measure of $N_3$, whereas orange squares use loss from $N_2$, as described in the text. The approximate noise floor from shot-to-shot number fluctuations is denoted by a shaded region.
\label{fig:spectrum}}
\end{figure}

For rapid single-shot measurement, the diffuse nature of $\Gamma_\mathrm{HFT}$ is disadvantageous. To avoid signal contamination by the residue of single-particle spin flips \cite{Navon:2025b} (see Fig.~\ref{fig:spectrum}), for a square pulse, one must probe at $\tilde\omega \gg (t_\mathrm{rf} \widetilde{C}/\tau_F)^{-2}$, where $t_\mathrm{rf}$ is the pulse time, $\tilde \omega \equiv \omega \tau_F$, and $\widetilde C \equiv C/N k_F$.
This constraint is more stringent than the conventional Fourier expectation $\omega \gg t_\mathrm{rf}^{-1}$ due to the $\omega^{-3/2}$ scaling of the HFT. Achieving a fractional transfer $\alpha$, which is $\Gamma t_\mathrm{rf}/N_2$ for the square pulse, requires rf power that scales as $(t_\mathrm{rf} \widetilde{C}/\tau_F)^{-4}\alpha$. The resource requirements can be improved with pulse shaping [see End Matter], yet practical limitations on $\Omega_{23}$ still constrain the feasibility of this approach for small $t_{\text{rf}}/\tau_F$.

Here, we demonstrate that rf-induced dimer projection can serve as a rapid and sensitive probe of the contact parameter $C$. This technique relies upon selecting a final state $\ket{3}$ with a positive scattering length $a_{13}>0$, ensuring the existence of a discrete Feshbach dimer resonance below the $13$ continuum \cite{Chin:2005cg,Ketterle:2008,Navon:2023}. 
Molecular conversion also underlies photo-excitation (PE), one of the earliest probes of the contact parameter \cite{Partridge:2005jh,Werner:2009vb,wang2021photoexcitation,Denschlag:2019,Pan:2022, Denschlag:2024a, Denschlag:2024b,Journeaux:2025}. Unlike PE, which involves bound-to-bound transitions, rf dimer projection is dominated by the open-channel component of the initial state.
We conjecture -- and confirm experimentally -- that the rf spectral weight of the dimer feature, $I_d$, is proportional to $C$ \footnote{In a different context, longitudinal field oscillation, Langmack {\it et al.}~\cite{Braaten15a} have shown that the contact operator couples interacting ensembles to ground-state dimer states.}. 
Using this method, we measure $C$ with a pulse duration shorter than $\tau_F$. In the scenario explored, the required rf power is proportional to $\widetilde C^{-1} ({t_\mathrm{rf}}/{\tau_F} )^{-2} \alpha$, offering a scaling more favorable to rapid measurement than the HFT approach.

The dimer feature is a major contributor to the clock shift 
$\Delta \equiv \int \! d\omega \, \Gamma\omega/ \int \! d\omega \, \Gamma$ of repulsive gases. In the weakly interacting regime, $\Delta$ corresponds to a mean-field shift of the peak transfer frequency, $\propto (a_{13}-a_{12})$. In the unitary Fermi gas (UFG), the peak shift remains finite even as $a_{12} \to \infty$, and at low temperature reveals a pairing gap \cite{Regal:2003ex,Gupta:2003,Ketterle:2008,Jin:2008,Mueller:2017,Jochim:2018,mukherjee2019,SpectrosocpyReview2021};
however, unlike the weakly interacting case, $\Delta$ has little to do with the location of the peak in $\Gamma$. 
Consider the clock shift in the zero-range case \cite{Baym:2007ki,Punk2007,Zhang:2008dc,Zhang:2009kq}
\begin{equation} \label{eq:clockshift} \Delta^\mathrm{zr} = \frac{-\hbar}{2 \pi m N a_{13} } \, C\,. \end{equation} 
For $a_{13}<0$, one can see that $\Delta^\mathrm{zr}$ is dominated by the HFT: the first moment of $\Gamma_\mathrm{HFT}^\mathrm{zr}$, integrated from some $\omega_\mathrm{min}$ to $\infty$, gives the expected (positive) $\Delta^\mathrm{zr}$ times $(2/\pi) \arctan\sqrt{\omega_a/\omega_\mathrm{min}}$, which approaches unity for $\omega_\mathrm{min} \ll \omega_a$. The surprising aspect of the rf clock shift is that it is determined not by the location of the peak in $\Gamma$, but by parts of the spectrum far from the peak, where $\Gamma$ is a small fraction of the peak transfer rate. Perhaps due to this challenge, $\Delta$ has yet to be measured in a UFG, and Eq.~\eqref{eq:clockshift} remains untested \footnote{Two Bose gas experiments have measured the two-body contact through a Ramsey clock shift \cite{Fletcher:2017,Beugnon:2021}. However, each of these operated in a long-interrogation-time limit where the accumulated phase measures the contact parameter through the shift in the resonance, rather than the first moment of $\Gamma$.}. 

Further consideration of the clock shift shows why $I_d \propto C$. Instead of the $a_{13}<0$ scenario, consider the rf spectrum with an equal and opposite $a_{13}>0$ (but the same $C$ and still in the zero-range limit). Eq.~\eqref{eq:clockshift} shows that the total $\Delta^\mathrm{zr}$ is now {\em negative}. The integral over the HFT is unaffected by the sign of $a_{13}$, such that its contribution is $|\Delta^\mathrm{zr}|$. If the only modification of the spectrum between the $a_{13}<0$ and $a_{13}>0$ cases is that the dimer feature appears (drawing its spectral weight from the single-particle residue) then {\em its first moment must be $2\Delta^\mathrm{zr}$}, to ensure the correct total shift. The implication for $I_d$ can be found in the limit where the feature width (comparable to the two-body continuum width) is much smaller than the binding energy $\hbar \omega_d = \hbar^2\kappa^2/m$, such that $\Delta_d = -I_d \omega_d$. 
In the z.r.\ case, $\omega_d = \omega_a$ and $\kappa = a_{13}^{-1}$, and thus $I_d^\mathrm{zr} = a_{13} C/(N \pi)$.

Generalizing this relationship beyond the z.r.\ case, we propose that 
\begin{equation} \label{eq:Id} 
I_d = \frac{\ell_d}{\pi}\frac{C}{N} \,, \end{equation} 
where $\ell_d$ is a $C$-independent length characterizing the strength of the dimer feature, including non-universal corrections.

The dimer-projection strength can be probed through a resonant $\omega=-\omega_d$ pulse of duration $t_\mathrm{rf} \ll 2 \pi \tau_F$ which results in a perturbative transfer fraction $\alpha_d = \Omega_d^2 t_\mathrm{rf}^2/4$. Here $\Omega_d$ is reduced from $\Omega_{23}$ by $I_d$: 
\begin{equation} \label{eq:RabiDI} 
\Omega_d = I_d^{1/2} \Omega_{23}\,, \end{equation} 
which is proportional to $C^{1/2}$. Similar scaling was found for PE dimer coupling in Ref.~\cite{Brantut:2021}. 
Equations~\eqref{eq:Id} and \eqref{eq:RabiDI} constitute new contact relations that link rf dimer projection to the wide range of other experimental observables discussed for instance in Refs.~\cite{Tan:2008a,Tan:2008b,Tan:2008c,Braaten:2008ez,Zhang:2009kq,Braaten:2012gh,Werner:2012um}. 

{\em Observed spectrum.} 
We prepare a spin-balanced mixture of $^{40}$K in equilibrium with typical $N = 3 \times 10^4$, $E_F/h = 15$\,kHz, and $\tau_F = 11\,\mu$s. The internal states {$\ket{1}$, $\ket{2}$, and $\ket{3}$} correspond to the three lowest hyperfine-Zeeman states of the ground-state manifold. Unless otherwise specified, the magnetic field is set to $B = 202.14(1)$\,G, such that $\ket{1}$ and $\ket{2}$ are at unitarity: $|a_{12}|^{-1} \ll k_F$. 
Populations $N_\sigma$ are measured by state-selective time-of-flight imaging. Figure~\ref{fig:spectrum}(a) depicts a typical rf transfer spectrum, including both the HFT and the dimer feature. We suppress signal contamination from the single-particle residue in HFT by using both long probe-pulse times ($t_\mathrm{rf} \sim 1\,$ms) and a Blackman pulse envelope.

Figure~\ref{fig:spectrum}(b) compares two measures of $\alpha$ across the HFT. 
Within the range $E_F \lesssim \omega \lesssim U_t$, where $U_t \sim 200$ kHz is the trap depth, $\alpha$ is measured by the ratio $N_3/(N_2+N_3)$. 
For larger $\omega$, the final states in the continuum are not trapped, so instead 
$\alpha$ is measured by $1 - N_2/\bar{N}_2^{\mathrm{ref}}$, where $\bar{N}_2^{\mathrm{ref}}$ is an average over reference images without rf drive. 
Above $U_t$, the latter protocol more accurately captures the $\omega^{-3/2}$ signature, although the signal-to-noise ratio suffers from shot-to-shot number fluctuation. 

The dimer feature is also observed through $\omega$- and $\Omega_{23}$-dependent loss, but where $\alpha_d$ is estimated as $(1 - N_2/\bar{N}_2^{\mathrm{ref}})/2$. The factor of $1/2$ in $\alpha_d$ accounts for an additional loss process arising from inelastic atom-molecule collisions \cite{quemener2012ultracold}: for every dimer produced, the constituent $\ket{1}$ and $\ket{2}$ atoms are lost, along with an additional $\ket{2}$ atom through this secondary collisional process \cite{SM}. 
We calibrate and correct for the systematic departure from linear response and finite loss due to $p$-wave interactions in the $13$ channel \cite{SM}.

\begin{figure}[tb!]
\centering
\includegraphics[width=\columnwidth]{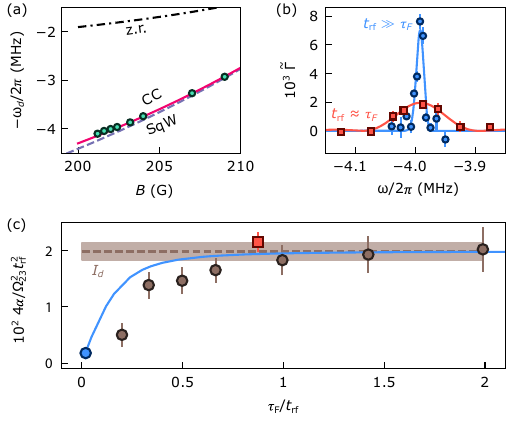}
\caption{
{\em Dimer feature.} 
{(a)} The rf detunings at which maximum dimer transfer is observed (points) are compared to various models of the binding energy (lines) across a range of magnetic fields $B$. 
{(b)} At $202\,$G, the lineshape $\widetilde \Gamma (\omega)$ of the dimer feature depends on the pulse time. 
For long pulses ($t_\mathrm{rf} = 640\,\mu\mathrm{s} \approx 50 \tau_F$, blue circles), the profile reveals the width of the continuum. The blue line has a width comparable to $E_F$, and serves as a guide to the eye. For short pulses ($t_\mathrm{rf} = 10\,\mu\mathrm{s} \approx 1.1 \tau_F$, red squares), the profile is determined by the rf pulse, and insensitive to $T$. The red line shows the $\mathrm{sinc}^2$ function described in the main text. 
{(c)} The fractional transfer fraction divided by $\Omega_{23}^2 t_\mathrm{rf}^2/4$ approaches a constant for short pulse times. In the short-pulse limit, this ratio gives the dimer weight $I_d$ through Eq.~\eqref{eq:RabiDI}. The data from (b) are highlighted as blue and red points. The blue line models the expected $I_d$ from convolving the long-pulse lineshape with $\mathrm{sinc}^2$ functions of various durations.} 
\label{fig:dimer}
\end{figure}

Figure \ref{fig:dimer}(a) shows the observed (negative) binding energy $-\hbar\omega_d$ versus various magnetic fields, taken as the peak detuning of the dimer feature response \footnote{The frequency of the peak in $\alpha_d$ is offset from the true dimer binding energy by a correction comparable to the collision energy \cite{Chin:2005cg}. We have not addressed this systematic, since its magnitude (roughly 1\%) is small compared to the measurement uncertainties in $I_d$ and $\Delta_d$.}. The observations deviate from the universal z.r.\ $\omega_d = \omega_a$, but agree with a coupled-channels (CC) calculation. A square-well (SqW) model \cite{SM}, which includes effective-range effects but not multi-channel effects, gives a dimer energy that is $\sim3\%$ larger than observed.

Figure~\ref{fig:dimer}(b) shows dimer transfer spectra with long ($t_\mathrm{rf} \gg \tau_F$) and short ($t_\mathrm{rf} \approx \tau_F$) pulses. The long-pulse spectrum, shown for $E_F/h=13\,$kHz and $T=0.3 T_F$ (where $T_F$ is $E_F/k_B$, and $k_B$ is the Boltzmann constant), has a width comparable to $E_F/\hbar$, which is consistent with thermal and interaction energy scales in the equilibrium UFG \cite{Braaten15a}. 
This structure is no longer resolved by the short-pulse spectrum; instead, the response is dominated by Fourier broadening. In linear response, a square pulse of length $t_\mathrm{rf}$ will result in $\alpha_d 
\propto (t_\mathrm{rf}/2 \pi) \mathrm{sinc}^2[ (\omega +\omega_d) t_\mathrm{rf}/2]$, whose FWHM is $\sim 2\pi/t_\mathrm{rf}$. This compares well to the observed response [see Fig.~\ref{fig:dimer}(b)]. 

Short resonant pulses have a transfer proportional to $\Omega_d^2$ and thus $I_d$, as described by Eq.~\eqref{eq:RabiDI}. We expect a single-shot measure of $I_d$ to be a valid probe of the contact for measurement times satisfying $2 \pi \tau_F \gg t_\mathrm{rf} \gg \omega_d^{-1}$ such that Fourier broadening dominates over the continuum width. Indeed, Figure~\ref{fig:dimer}(c) shows that this condition is satisfied for $\tau_F$-scale measurements where $\tau_F/t_\mathrm{rf} \gtrsim 1$, and $\alpha_d/t_\mathrm{rf}^2$ becomes independent of pulse time. 

\begin{figure}[tb!]
\centering
\includegraphics[width=\columnwidth]{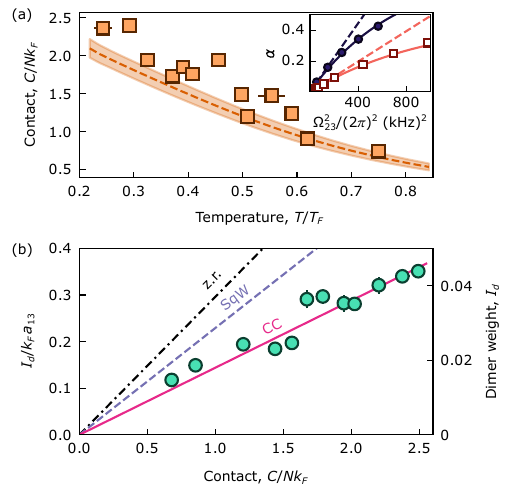}
\caption{
{\em Dimer spectral weight versus contact.} 
(a) $\widetilde C$ (orange squares) measured at a single frequency in the HFT is shown at various $T/T_F$. 
A calculation of the harmonic-trap-averaged contact is shown as a dashed orange line, with a shaded band representing the systematic uncertainty of thermometry. Inset:\ examples of the linear-response calibration for $T/T_F \approx 0.3$ (upper) and $T/T_F \approx 0.6$ (lower). 
(b) Single-frequency measurements of the dimer spectral weight (green circles) are shown versus measured $\widetilde C$ from HFT. Data are compared to three model predictions: the universal z.r.\ limit (dash-dotted black), the SqW model with a finite effective range (dashed purple), and a molecular CC calculation (solid magenta). 
\label{fig:spectralweight}} 
\end{figure}

{\em Proportionality with Contact.} 
To verify $I_d$ as a probe of $C$, we study unitary clouds at a wide range of temperatures. Figure~\ref{fig:spectralweight}(a) shows $C/Nk_F$, measured by probing $\Gamma_\mathrm{HFT}$ with 200-$\mu$s-long Blackman pulses at $\omega=2 \pi \times 100\,$kHz, versus $T/T_F$. 
We assume that finite-range and multichannel effects are significant only at $\omega \gtrsim \omega_a$, 
such that Eq.~\eqref{eq:GammaAsymptotic} remains accurate at this $\omega/\omega_a \sim 0.05$. 
The entrance channel will have a closed-channel fraction proportional to $C/N$, but only up to $0.8\%$ for these conditions. 
The rf frequency is chosen to be less than $U_t/\hbar$ [see Fig.~\ref{fig:spectrum}(b)]. 
Data in Fig.~\ref{fig:spectralweight}(a) are comparable to prior measurements under similar conditions \cite{Hoinka:2013dx,Sagi:2018} and follow the trend anticipated by a Luttinger-Ward calculation of the contact averaged over the in-trap density profile with no free parameters [see End Matter]. The measured $\widetilde C$ 
are $16(4)\%$ larger than anticipated by this model, comparable to systematic uncertainties in number of atoms and temperature. 

Figure~\ref{fig:spectralweight}(b) shows a model-independent comparison of the dimer weight and $\widetilde C$ found from the HFT, for unitary Fermi gases prepared at various $T/T_F$. A single point represents the average of about 30 datasets, each composed of four experiment cycles
that interleave measurements of $C$ from the HFT and $I_d$ from the dimer, along with zero-drive reference measures for both. 
Probe conditions are chosen to remain close to the linear-response regime, and we compensate for nonlinear effects by correcting $\alpha$ to an extrapolated linear-in-$\Omega_{23}^2$ regime, as seen in the inset of Fig.~\ref{fig:spectralweight}(a) for the HFT. This calibration is data-intensive, as it depends both on $\omega$ and on $T/T_F$. Figure~\ref{fig:spectralweight}(b) reports $I_d/k_F a_{13}$ in order to account for the variation in $E_F$ between datasets; the unscaled dimer weight, plotted on the right axis, are calculated using an average $k_F$. 
We find a clear proportionality between $I_d$ and $C$. Interpreted with Eq.~\eqref{eq:Id}, a best-fit $\ell_d$ is found to be $100(3)(8)\,a_0$, with uncertainties that are respectively statistical and systematic.

Data in Fig.~\ref{fig:spectralweight}(b) are compared to matrix elements anticipated by the z.r., SqW, and CC models. 
Clear deviation from the z.r.\ $\ell_d = a_{13} \approx 222\,a_0$ indicates that $^{40}$K is not in the universal-dimer limit, as already anticipated by $\kappa \neq a_{13}^{-1}$ in Fig.~\ref{fig:dimer}(a), since $\kappa$ governs the exponential decay of the dimer state beyond the range of the potential.
A $30\%$ correction can be attributed to finite-range effects, which modify the matrix element: 
the SqW model \cite{SM} finds 
$\ell_d \approx 160\,a_0$. 
The CC matrix element reveals a further correction due to multi-channel effects, and yields $\ell_d \approx 100.7\,a_0$, in agreement with the measured dimer strengths. Using the CC value for $\kappa$, the deviation of our best-fit line from the z.r. case is given by $\kappa \ell_d = 0.67(2)(6)$.

{\em Clock shift.} Our broad-spectrum measurements of $\Gamma(\omega)$ enable us to constrain all contributors to the $\Delta$ of a UFG with $\widetilde{\Gamma} \gtrsim 10^{-5}$. 
Table~\ref{tab:clock} summarizes $\Delta$ from the dimer, HFT, and near-resonant sectors of the rf spectrum, for the example case of $\widetilde C = 2.2$. Also shown are predictions from various models \footnote{A full-spectrum coupled-channels calculation of $\Delta_\mathrm{HFT}$ and total $\Delta$ is beyond the scope of this work.}.
The deviations from z.r.\ predictions can be expressed using the dimensionless ratios $A_d \equiv \Delta_{d}/\Delta_{d}^\mathrm{zr}$, $A_\mathrm{HFT} \equiv \Delta_\mathrm{HFT}/\Delta_\mathrm{HFT}^\mathrm{zr}$, and $A_c \equiv \Delta/\Delta^\mathrm{zr}$. If the HFT and dimer were the only contributors to the clock shift, then the relation between these factors would be $A_c = 2 A_d - A_\mathrm{HFT}$. 

The largest observed contribution in Tab.~\ref{tab:clock} is from the dimer. We also note that $\Delta_d$ is consistent with both the CC prediction and the z.r.\ prediction. Since $A_d = I_d \omega_d/(2 |\Delta^\mathrm{zr}|) = \kappa^2 \ell_d a_{13}$, the agreement with $\Delta_{d}^\mathrm{zr}$ involves a coincidental cancellation of the non-universal corrections to both $\omega_d \propto \kappa^2$ and $I_d \propto \ell_d$. 

The HFT contribution $\Delta_\mathrm{HFT}$ can only be bounded by our data due to the noise floor of $\widetilde{\Gamma}$ [see Fig.~\ref{fig:spectrum}(b)]: using only with $\widetilde{\Gamma} \gtrsim 10^{-5}$, we find $A_\mathrm{HFT} > 0.5$. A similar bound on $\Delta_\mathrm{HFT}$ is drawn from the observed spectrum in Ref.~\cite{Sagi:2018}. 
We anticipate that $\Delta_\mathrm{HFT}$ will be sensitive to any effects that modify Eq.~\eqref{eq:GammaAsymptotic} at $\omega \gtrsim \omega_a$ \cite{SM}. 

We remark that the structure of the near-resonant ($|\omega| \lesssim 3 E_F/\hbar$) spectral feature [see inset of Fig.~\ref{fig:spectrum}(a)] does not provide a leading-order contribution. We estimate a positive shift not more than $\approx 0.4 E_F/\hbar$, which is consistent with more detailed studies of peak shifts \cite{SpectrosocpyReview2021}. That this shift is an order of magnitude smaller than other contributions (and their sum) in Tab.~\ref{tab:clock} illustrates why studies of the near-resonant peak in $\Gamma$ do not constrain $\Delta$. 

The sum of the observed contributions determines a lower bound to $\Delta$, here $\approx -8 E_F/\hbar$. In dimensionless form, we find that $A_c < 1.5$. This upper bound to $|\Delta|$ already excludes the SqW prediction ($A_c \approx 2.1$), despite its accurate prediction of dimer binding energies [Fig.~\ref{fig:dimer}(a)]. Table \ref{tab:clock} also shows that the observed bound on the total clock shift excludes a sum-rule calculation ($A_c \approx 2.6$) based on \cite{Baym:2007ki,Punk2007} with an effective-range correction \cite{SM}. The spectral clock shift of unitary $^{40}$K must therefore depend on corrections beyond the realm of low-energy scattering, calling into question its utility as a measure of $C$. In contrast, high-energy contributions did not seem to affect interferometric measurements of $C$ in Bose gases \cite{Fletcher:2017,Beugnon:2021}, in which the time between Ramsey pulses effectively creates a low-pass filter. 

\begin{table}[t!]
\caption{Clock shift at $C = 2.2 N k_F$, measured with a gas of $E_F/h \approx 14.1\,$kHz. The observed contribution from the HFT includes the observed $\Gamma$ from $50\,$kHz up to $2.9\,$MHz, and thus sets a lower bound. 
\label{tab:clock}}
\centering
\begin{ruledtabular}
\begin{tabular}{cccc} 
& \multicolumn{2}{c}{$\hbar \Delta/E_F$} &  \\
contribution & measured & predicted & model \\ \hline
dimer & $-11.3(8)$ & $-11.3$ & z.r. \\
 &  & $-18.4$ & SqW \\
 &  & $-11.3$ & CC \\  [0.5ex]
HFT & $> +2.6(1)$ & $+5.7$ & z.r.  \\
& & $+6.6$ & SqW \\ [0.5ex]
near-resonant & $ \lesssim +0.4$ & & \\ [1ex]
total & $> -8.3(8)$ & $-5.7$ & z.r.  \\ 
 & & $-11.8$ & SqW \\
 &  &  $-15(1)$ & sum rule
\end{tabular}
\end{ruledtabular}
\end{table}

{\em Discussion and conclusion.} 
Radio-frequency dimer projection can be used in any system with $a_{13}>0$. We anticipate that for a UFG of $^6$Li near its 690\,G Feshbach resonance, $\ell_d$ would approach the universal case, since its resonant closed-channel fraction is smaller than $^{40}$K, as is the relative magnitude of the effective range. 
Estimates \cite{SM} of $A_\mathrm{HFT}$, $A_c$, and $A_d$ indicate they should be within 10\% of the zero-range limit. For typical experimental conditions, dimer projection should also provide a route for $\tau_F$-scale measurements of contact dynamics in $^6$Li. 

A natural extension of our work is to generalize the dimer-contact relations beyond unitarity, i.e., to $a_{12}^{-1} \neq 0$, enabling exploration of rapid contact dynamics across the BEC-BCS crossover. Preliminary evidence that $I_d$ can probe weaker correlations on the BCS side is suggested by the increased drive strengths we find to be required when measuring $-\omega_d$ for Fig.~\ref{fig:dimer}(a). 
Other promising directions include the adaptation of rf dimer projection to probe higher-partial-wave correlations, such as the $p$-wave contact, and to two-dimensional systems where dimer states also exist for $a_{13}<0$. 

In summary, we introduce rf dimer projection as a new tool for probing contact dynamics in ultracold gases. Characterization of the dimer feature enabled significant progress towards measuring the clock shift in a unitary Fermi gas, although its complete determination, both experimentally and theoretically, remains a challenge for future studies. 
Dimer projection provides a measure of the contact parameter in a time shorter than $\hbar/E_F$, due to the gapped resonant frequency and high spectral intensity of the dimer feature. This capability opens new avenues for studying nonequilibrium pair-correlation dynamics in strongly interacting systems. 


\begin{acknowledgments} We thank K.\ G.\ Jackson for early work on this project, Songtao Huang and N.\ Navon for detailed discussions and sharing unpublished measurements of the dimer feature in $^6$Li. 
We thank F.\ Chevy, R.\ Fletcher, L.\ LeBlanc, A.\ Vutha, Shizhong Zhang, and M.\ Zwierlein for stimulating conversations. This research is supported by NSERC, by AFOSR FA9550-24-1-0331, by Deutsche Forschungsgemeinschaft (German Research Foundation), Project No.\
273811115 (SFB1225 ISOQUANT) and under Germany’s Excellence Strategy EXC2181/1-390900948 (the Heidelberg STRUCTURES Excellence Cluster), and by the National Natural Science Foundation of China Grant No.~12474270. The data that support the findings of this article are openly available \cite{DataBorealis}. 
\end{acknowledgments}

\bibliography{bibSumRule}

\onecolumngrid 
\begin{center}
\vspace{\columnsep}
\textbf{End Matter}
\vspace{0.25\columnsep}
\end{center}
\twocolumngrid

{\em Relative power efficiency of HFT and dimer probes of contact.} Short-pulse rf probes of contact must be chosen such that they do not excite the single-particle residue, which does not respond proportionally to $C$. In the $\omega^{-3/2}$ regime of the HFT, the fractional transfer is
\begin{equation} \label{eq:HFTresponse}
\alpha_\mathrm{HFT} = 2^{-3/2} \pi^{-1} \tau_F t_\mathrm{rf} \Omega_{23}^2 \widetilde C \tilde\omega^{-3/2}\,. \end{equation}
For a uniform amplitude rf probe, the resonant response is driven by an off-resonant transfer probability proportional to the side bands of the square pulse Fourier transform
\begin{equation} \alpha_\mathrm{res} \approx 2^{-1} \pi t_\mathrm{rf} \Omega_{23}^2 \delta_t(\omega)\,, \end{equation} 
where $\delta_t(\omega) = (t_\mathrm{rf}/2 \pi) \mathrm{sinc}^2(t_\mathrm{rf} \omega/2)$. This is a slight overestimate, since it neglects the depletion of the single-particle residue due to correlations. 
For large detuning, $\omega t_\mathrm{rf} \gg 1$, then $\delta_t(\omega) \approx 1/\pi \omega^2 t_\mathrm{rf}$, where we have taken the average value $\sin^2(\omega t_\mathrm{rf}/2) = 1/2$. Then 
$\bar\alpha_\mathrm{res} \approx {\Omega_{23}^2}/({2\omega^2})$. 
The ratio of responses is 
\begin{equation} \frac{\alpha_\mathrm{HFT}}{\bar\alpha_\mathrm{res}}
\approx 
\frac{ t_\mathrm{rf} \widetilde C \tilde\omega^{1/2} }{2^{1/2} \pi \tau_F} \,.\end{equation} 
Let us set this ratio equal to $R$, and require that $R \gg 1$ for a faithful contact probe. The detuning must then be 
\begin{equation} \label{eq:cleanprobe}
\tilde\omega \approx R^2 \frac{2 \pi^2 \tau_F^2}{t_\mathrm{rf}^2 \widetilde{C}^2} \,. \end{equation} 
It is surprising at first that the required detuning scales as the pulse length squared; the typical Fourier argument (that $\omega$ need only scale as $1/t_\mathrm{rf}$) breaks down because of the $\omega^{-3/2}$ reduction of the HFT response. For our typical $E_F/h \sim 15\,$kHz, the roll-over to $\omega^{-5/2}$ scaling will need to be included for $R\gtrsim 2 \widetilde{C}$. 

Now we can consider the rf power required for response $\alpha$ in the short-pulse regime $t_\mathrm{rf} \lesssim \tau_F$, while respecting Eq.~\eqref{eq:cleanprobe}.
One finds
\begin{equation}\label{eq:square_pulse_power_req}
\frac{\Omega^2_{23}}{\alpha_\mathrm{HFT}}\bigg |_{\mathrm{Square}}\approx \frac{8\pi^4 R^3}{\tau_F^2 \widetilde{C}^4} \left(\frac{t_\mathrm{rf}}{\tau_F}\right)^{-4}\,.
\end{equation}

Improved power-efficiency can be gained by using a Blackman pulse shape, which suppresses frequency side bands. A similar calculation to the above with a Blackman Fourier transform (of full length $t_\mathrm{rf}$) gives $R>10^3$ when $\tilde \omega \gtrsim 6\pi \tau_F/t_\mathrm{rf}$, which is accessible within the $\tilde\omega^{-3/2}$ regime of the HFT. Even in the $t_\mathrm{rf} \gg \tau_F$ regime, the permissive detuning constraint of the Blackman-windowed pulse enables larger response while maintaining pure signal. The Blackman window is used for the measurements in Figs.~\ref{fig:spectrum} and \ref{fig:spectralweight}(a) for this purpose. By measuring transfer fraction for negative detunings with both square- and Blackman-shaped pulses, we estimate our Blackman window suppresses sidebands by more than a factor of 30.

Respecting the signal purity condition for the detuning of the Blackman-windowed pulse, we consider the rf power required for response $\alpha$ in the short-pulse regime. One finds 
\begin{equation} \label{eq:HFTpower} 
\frac{\Omega_{23}^2}{\alpha_\mathrm{HFT}} \bigg |_{\mathrm{Blackman}} = \frac{\mathcal{N}}{\tau_F^2\widetilde{C}}\left(\frac{t_\mathrm{rf}}{\tau_F} \right)^{-5/2} \,, \end{equation} 
where the numerical factor $\mathcal{N} \approx 2.4 \times 10^3$. 

By comparison, the dimer response is $\alpha_d = I_d \Omega_{23}^2 t_\mathrm{rf}^2/4$, such that with Eq.~\eqref{eq:Id}, 
\begin{equation} \label{eq:Dpower} 
\frac{\Omega_{23}^2}{\alpha_d} \bigg |_{\mathrm{Square}}= \frac{4\pi}{k_F \ell_d \tau_F^2 \widetilde{C}} \left(\frac{t_\mathrm{rf}}{\tau_F} \right)^{-2}\,. \end{equation}
With $E_F/h=15\,$kHz and the measured $\ell_d = 10^2\,a_0$, then $4\pi/k_F \ell_d \sim 10^2$.

Since the detuning of the dimer is fixed at $-\omega_d$, insisting on a purity $\alpha_d / \bar\alpha_\mathrm{res} \geq R$ results in a constraint on the pulse time: 
\begin{equation} \label{eq:dimerpulselength}
t_\mathrm{rf} \gtrsim \omega_d^{-1} \sqrt{ \frac{2 \pi R}{\ell_d k_F \widetilde{C}} }\,. \end{equation}
For $\omega_d = 2 \pi \times 4\,$MHz, $\ell_d \approx 10^2\,a_0$, $\widetilde C \approx 1$, and $R=10$, one finds $t_\mathrm{rf} \gtrsim 1\,\mu$s. 

Comparing the power requirements at $t_\mathrm{rf} = \tau_F$ and assuming that Eq.~\eqref{eq:dimerpulselength} remains satisfied, then the ratio of powers given by Eqs.~\eqref{eq:HFTpower} and \eqref{eq:Dpower} is 
\begin{equation} 
\frac{\Omega^2_{23,\mathrm{HFT}}}{\Omega^2_{23,\mathrm{d}}} = 
\frac{\mathcal{N} k_F \ell_d}{4\pi} \,,\end{equation}
which is $\approx 24$ for our typical $k_F$. In other words, the power requirement to probe at $t_\mathrm{rf} \approx \tau_F$ is reduced by over one order of magnitude using the dimer-projection approach. 

{\em Harmonic trap-averaged contact.} In Fig.~\ref{fig:spectralweight} we compare the measured, trap-integrated contact with theoretical predictions. Here we describe how these are obtained: We start with Luttinger-Ward calculations for the contact density $\mathcal C$ of a uniform UFG and then integrate over the inhomogeneous density profile in the trap. 

The Luttinger-Ward calculation \cite{Haussmann:2009, Enss:2011gk} is based on a two-channel model in the zero-range limit, where the contact is determined from the density of dimers, or equivalently from the large-momentum tail of the fermion momentum distribution. The dimer Green function is computed via the T-matrix, which captures repeated scattering between two fermions. In the Luttinger-Ward approach the T-matrix is computed self-consistently using dressed Green functions for the fermions in the medium, which acquire a finite lifetime due to scattering off (virtual) dimers; for details on the implementation see \cite{Haussmann:2009, Enss:2011gk, enss2019bulk}.  This computation yields the local contact density $\mathcal C = sk_F^4 = s(3\pi^2n)^{4/3}$, where both the dimensionless contact $s(\mu)$ and the density equation of state $n(\mu)$ depend on the chemical potential $\mu$ and temperature.

In order to compute the trap-integrated contact $C$, we first determine the density profile of the trapped unitary gas. Within the local density approximation (LDA) it follows as $n(r) = n[\mu(r)]$ with the equation of state $n(\mu)$ evaluated at the local chemical potential $\mu(r)=\mu-V(r)$ for a given external trapping potential $V(r)$. We employ the measured equation of state \cite{Mark2012} and construct an all-temperature fitting function that is represented as a barycentric rational approximant, complemented by the virial expansion for high temperatures.  For given experimental values of particle number $N$, temperature $T$ and potential $V(r)$, we adjust the global chemical potential $\mu$ in the density profile $n(r) = n[\mu-V(r)]$ until it integrates to the desired particle number $N=\int d^3r\,n(r)$.  Next, the dimensionless contact $s(\mu)$ from the Luttinger-Ward calculation is also encoded into a global fitting function. Then, the trap-integrated contact is obtained as an integral over the trap volume at fixed temperature,
\begin{equation}
    C = \int d^3r\, \mathcal C(r) = \int d^3r\, s[\mu(r)]\, \{ 3\pi^2n[\mu(r)] \} ^{4/3}\,,
\end{equation}
with position-dependent local chemical potential $\mu(r)=\mu-V(r)$. The intensive measure for the contact, $C/Nk_F$, is shown in Fig.~\ref{fig:spectralweight}. As a model-independent reference scale, we use the global Fermi momentum of an ideal Fermi gas at zero temperature, as given by the relation $E_F = \hbar\Bar\omega_\text{trap} (3N)^{1/3} = \hbar^2 k_F^2/2m$.

{\em Multichannel wavefunctions.} We use standard coupled channels calculations \footnote{J. M. Hutson and C. R. Le Sueur, BOUND: a program for bound states of interacting pairs of atoms and molecules \cite{Hutson2019a}, Version 2022.0, and MOLSCAT: A program for non-reactive quantum scattering calculations on atomic and molecular collisions ~\cite{Hutson2019b}, Version 2020.01} to calculate bound state and scattering properties based on the known matrix elements of the full spin Hamiltonian and the singlet and triplet molecular potentials for the $^{40}\mathrm K_2$ dimer molecule \cite{Falke2008}.  The Fermi gas is a spin mixture of $\ket{1}$ and $\ket{2}$ atoms, which interact in an $s$-wave with a $M_\mathrm{tot} = -8$ projection of the total angular momentum. The closed channel that causes the $s$-wave Feshbach resonance is primarily $\{1,18\}$, where state 18 is the highest energy level, adiabatically connected to the $\ket{f=7/2,m_f=-7/2}$ atomic level at $B=0$. The CC calculation finds the resonance pole at $B_0=202.11\,$G with an on-resonance effective range of $115.3\,a_0$; prior comparison to data corrects this value by $+0.04(2)\,$G \cite{AhmedBraun:2021}, so that our laboratory field, 202.14(1)\,G, is within uncertainty of unitarity. 

Radiofrequency spectroscopy rotates atoms from $\ket{2}$ to $\ket{3}$, such that interactions occur with $M_\mathrm{tot} = -7$. At $B_0$, the scattering length is $a_{13} = 221.5\,a_0$, and the effective range is $r_{e,13} = 103.5\,a_0$. Scattering is slightly enhanced by the $s$-wave Feshbach resonance at $224$\,G above the background scattering length of $167\,a_0$. The effective ranges are dominated by the van der Waals contribution, $r_{e,0} (1 - 2 \bar a/a_{13} + 2 \bar a^2/a_{13}^2)$, which gives $r_{e,13} \approx 108\,a_0$ with $\bar{a} \approx 62.2\,a_0$ and $r_{e,0} = \Gamma_{1/4}^4 \bar a /(6 \pi^2) \approx 181\,a_0$. While the zero-range phase shift would anticipate that $\eta \to -\pi/2$ at infinite collisional energy, a finite range causes the phase to cross through $-\pi/2$ at $E_k = \hbar^2/(2 m r_{e,13} a_{13})$, or about $E_k/h \approx 7.9\,$MHz here. Near and beyond this energy, the effective-range description qualitatively fails to predict the CC scattering phase. 

The dimer state at $B_0$ is bound by $h \times 4.019$\,MHz. Projected onto separated-atom spin states, it is primarily $\{1,3\}$ (93\%) but with significant $\{1,17\}$ (3.9\%) and $\{2,18\}$ (2.9\%), where state $17$ has $\ket{7/2, -5/2}$ atomic character. Transverse field oscillation couples scattering states into the dimer state via $-\bm{\hat\mu} \cdot \bm{B_\mathrm{rf}}(t)$, which acts both on the nuclear and electronic magnetic moment as $g_S \mu_B \hat{S}_x B_\mathrm{\perp}(t)$ and $g_I \mu_B \hat{I}_x B_\mathrm{\perp}(t)$ respectively, where $\mu_B$ is the Bohr magneton and $B_\perp$ is the component of the oscillating field that is perpendicular to the static Feshbach field. However, since $g_S \approx 2.0023$ and $g_I = 0.000176$, the electronic moments dominate the matrix element such that inclusion of the nuclear spin-flip revises $\ell_d$ by only $\approx 0.5\%$. 

The transition strength thus includes both spin-flip and spatial matrix elements: 
\begin{equation} \bra{\Psi_d} \hat{S}_+ \ket{\Psi_\mathrm{en}} = 
\sum_{\alpha,\beta} \bra{u_\alpha}\ket{u_\beta} \bra{\alpha} \hat{S}_+ \ket{\beta}\,,
\end{equation}
where $\alpha$ denotes the $\{1,3\}$, $\{1,17\}$, and $\{2,18\}$ dimer spin components, and $\beta$ denotes the $\{1,2\}$ and $\{1,18\}$ scattering spin components. Since some of the signs of these elements are negative, the full matrix element is roughly 15\% less than the leading term $\bra{u_{13}}\ket{u_{12}} \bra{13} \hat{S}_+ \ket{12}$.

The relation between $I_d = |\bra{\Psi_d} \hat{S}_+ \ket{\Psi_\mathrm{en}}|^2$ and $C$ was evaluated with a set of entrance-channel wave functions that were artificially confined to a spherically symmetric harmonic oscillator (SHO) potential. The long-wavelength details of the wavefunctions were unimportant, since $I_d$ and the contact parameter depend only on the short-range ($r \lesssim \kappa^{-1}$) properties of the wavefunctions. Here, $\kappa^{-1} \approx 150\,a_0$, while the harmonic oscillator lengths $a_\mathrm{ho}$ where $\sim 10^{4} a_0$. The first eigenstate of a SHO at unitarity has a contact $C/N = 4 \pi^{1/2} a_\mathrm{ho}^{-1}$. (Here $N=2$, and we report the intensive $C/N$ to compare to the many-body case.) The $C$ for each $\Psi_\mathrm{en}$ was confirmed by comparison to $C/N = 2 \pi f_\mathrm{cl}/R_*$, where $f_\mathrm{cl}$ is the closed-channel fraction and $R_* = \hbar^2 (m a_\mathrm{bg} \mu_b \Delta B)^{-1}$. For the $12$ resonance used here, $a_\mathrm{bg}=167\,a_0$, $\mu_b \approx 1.68 \mu_B$, and $\Delta B = 6.9\,$G, such that $R_* = 33\,a_0$. Using a range of SHO strengths from $2\,$kHz to $16$\,kHz, $I_d$ was found to be linear in $C/N$ with a slope $\ell_d/\pi$, with $\ell_d = 100.7\,a_0$, such that $\kappa \ell_d = 0.672$. When scaled by $k_F$, the range of contacts for these wavefunctions correspond to $\widetilde C \approx 1.3$ to $\widetilde C \approx 3.7$, comparable to those measured by the experiment. 

\nocite{schaefer2010dissipative,cao2011universal,Ku_2012,VanHoucke2012,Jochim:2013}

\ifsm

\clearpage
\appendixpageoff
\appendixtitleoff
\begin{appendices}

\begin{center}
\huge
Supplemental Material
\normalsize
\end{center}

\setcounter{equation}{0}
\setcounter{figure}{0}
\setcounter{table}{0}
\setcounter{page}{1}
\makeatletter
\renewcommand{\theequation}{S\arabic{equation}}
\renewcommand{\thefigure}{S\arabic{figure}}
\renewcommand{\thesection}{S\arabic{section}}

\section{Square-Well Model} 

The spherical square-well (SqW) model uses an interatomic potential $V_\sigma(r) = -V_\sigma$ for $r<r_0$, and $0$ outside of it. We consider the channels $\sigma = i,f$, i.e., $\sigma=12$ for the initial channel and $\sigma=13$ for the final channel. We will use the same $r_0$ for the two channels. A long-range boundary condition is a hard wall at radius $L \gg r_0$. The positive-energy eigenstates of the relative wave functions $\phi_\sigma$ can be indexed by a wave number $k_\sigma$, satisfying $k_\sigma L + \delta_\sigma(k) = n \pi$ 
for integer $n$, where $\delta_\sigma$ is the scattering phase shift of the $r>r_0$ wave function. The wave functions are (for $E>0$) 
\begin{equation} \begin{aligned} \label{eq:SqWphi} 
\phi_{\sigma,>} &= \mathcal{N}_{\sigma,>} \frac{\sin(k_\sigma r + \delta_\sigma)}{r}\, , \quad\mbox{for}\quad r>r_0\,, \\
\phi_{\sigma,<} &= \mathcal{N}_{\sigma,<} \frac{\sin(s_\sigma \tilde{r})}{r}\, , \quad\mbox{for}\quad r<r_0\,, 
\end{aligned} \end{equation} 
where $\eta_\sigma \equiv (m V_\sigma r_0^2)^{1/2} $ is the scaled momentum associated with the well depth, and $s_\sigma \equiv (\eta_\sigma^2 + \tilde k_\sigma^2)^{1/2}$. (We set $\hbar=1$ for this section.) Variables are made dimensionless using the range $r_0$: $\tilde r \equiv r/r_0$, $\tilde k_\sigma \equiv r_0 k_\sigma$. We consider $\eta_\sigma\gg1$ such that there are a large number of dimer states held by the square well potentials, similar to the situation due to the realistic interatomic interactions. 

The continuity of $\phi(r)$ and $d\phi(r)/dr$ at $r=r_0$ connects the normalization factors, $\mathcal{N}_{\sigma,>}$ and $\mathcal{N}_{\sigma,<}$, and determines the phase shift via 
\begin{equation} \label{eq:SqWBC} s_\sigma \cot{s_\sigma} = \tilde k_\sigma \cot{(\tilde k_\sigma + \delta_\sigma)}\,,\end{equation} 
whose solution is 
\begin{equation} \label{eq:SqWphase}
\cot \delta_\sigma = - \frac{s_\sigma + \tilde k_\sigma \tan{\tilde k_\sigma} \, \tan{s_\sigma}}{s_\sigma \tan{ \tilde k_\sigma} - \tilde k_\sigma \tan{s_\sigma}}\,. \end{equation}
The low-energy expansion of this phase shift can determine the scattering length $a$ and effective range $r_e$, by matching with $\cot \delta = -1/ka + r_e k/2$: 
\begin{equation} \label{eq:SqWaS}
\tilde{a}_\sigma \equiv \frac{a_\sigma}{r_0} = 1 - \tan{(\eta_\sigma)}/\eta_\sigma\,, \end{equation}
and 
\begin{equation} \label{eq:SqWre}
\tilde{r}_{e,\sigma} \equiv \frac{r_{e,\sigma}}{r_0} = 1 - \frac{1}{3 \tilde{a}_\sigma^2} - \frac{1}{\eta_\sigma^2 \tilde{a}_\sigma}\,. \end{equation}
The unitary case occurs in the $\sigma$ channel when $\eta_\sigma=\eta_j \equiv (j + \frac12) \pi$. The dimer state in the $13$ channel projected to in our experiment evolves continuously from the Feshbach dimer generated at $B=224.2$ G; $\eta_{13}$ shall be slightly larger than a unitary value $\eta_j$ by $\delta \eta = x/\eta_j$. In such a ``fine-tuned'' case that $\eta_{13}$ is near the same $j$ resonance as the $12$ channel, 
$\eta_{13}^2 - \eta_{12}^2 = 2 x + x^2/\eta_j^2 \approx 2x$. One can show that in the limit $\eta_j \gg 1$ and $\delta \eta \ll 1$, 
\begin{equation} \label{eq:finetuning}  
\eta_{13}^2 - \eta_{12}^2 \approx \frac{2}{\tilde a_{13} - 1}\,,\end{equation}
or $\tilde a_{13} - 1 \approx x^{-1}$.

For the $2\to3$ spectroscopy of $^{40}$K at the $12$ unitary point, we choose the parameters $r_0 = r_{e,12}=115\,a_0$, $\eta_{12}=\eta_j$, $\eta_{13} = \eta_j + x/\eta_j$ with $x=1.080$. This choice yields $r_{e,13} = 104.7\,a_0$ and $a_{13} = 221.5\,a_0$. We choose $j=100$; however so long as $x=\eta_j \delta \eta$ is fixed, these final-state parameters are also found for any $j \gg 1$. 

\begin{table}[b!]
\caption{{\bf Scattering parameters} found from coupled-channels calculations for $^{40}$K at the resonant field for the $12$ channel, compared to those used in the square-well model. 
\label{tab:model}}
\centering
\begin{ruledtabular}
\begin{tabular}{ccc} 
~~~~~~~~~~~~~ &  CC & SqW \\ \hline
$a_{12}$ & $\mathrm{pole}$ & $\mathrm{pole}$ \\
$r_{e,12}$ & $ 115.3 \, a_0$ & $ 115.0 \, a_0$ \\
$a_{13}$ & $ 221.5 \, a_0$ & $ 221.5 \, a_0$  \\
$r_{e,13}$ & $ 103.5  \, a_0$ &  $104.7 \, a_0$  
\end{tabular}
\end{ruledtabular}
\end{table}

The contact parameter in the $12$ channel for this model is determined by the pair probability density at $r=r_0$: 
\begin{equation} \label{eq:CfromNp3} 
C_\mathrm{SqW} \equiv |4 \pi r_0 \phi_{12,>}(r_0)|^2 = (4 \pi)^2 \mathcal{N}_{12,>}^2 \sin^2(\tilde{k}_{12}+ \delta_{12})\,. \end{equation} 
Considering the zero-energy limit, $\tilde k_{12} \to 0$, at unitarity, 
\begin{equation} \label{eq:CSqWU}
C_\mathrm{SqW,U} = 16 \pi^2 \mathcal{N}^2\,, \end{equation} 
where $\mathcal{N}=\mathcal{N}_<=\mathcal{N}_>$ in the unitary case. Since here $N=2$, we have $C_\mathrm{SqW,U}/N = 8 \pi^2 \mathcal{N}^2$. For sufficiently large $L$, one finds $\mathcal{N}_{>}^2 \approx (2 \pi L)^{-1}$ and correspondingly $C_\mathrm{SqW,U}/N \approx 4 \pi/L$. (For brevity, we will simply write $C$ for the remainder of this section.)

The square-well model supports a finite set of negative-energy solutions $E = -\hbar^2 \kappa^2 /m$ in the $13$ channel, in addition to the positive-energy solutions. Instead of Eq.~\eqref{eq:SqWphi}, the dimer has an outer wave-function 
\begin{equation} \label{eq:phidSqW}
\phi_d = \mathcal{N}_d \exp(-\kappa r)/r \qquad \mbox{for} \quad r>r_0\,, \end{equation}
where $\kappa$ is found through connecting with the wave-function for $r<r_0$. Instead of Eq.~\eqref{eq:SqWphase}, we have
\begin{equation} \tilde\kappa = - \sqrt{\eta_{13}^2 - \tilde\kappa^2} \cot \sqrt{\eta_{13}^2 - \tilde\kappa^2}\,, \end{equation}
which has multiple solutions. In the case $\eta_{13} = \eta_j + x/\eta_j$ with $\eta_j\gg1$ and $x/\eta_j\ll1$, and we are looking for the least bound dimer of $\tilde \kappa\sim O(1)$, one can show 
\begin{equation} \tilde \kappa \approx x - \left(\frac12 + \frac{x}{\eta_j^2}\right) \tilde \kappa^2\,, \end{equation}
which gives in the leading order
\begin{equation} \label{eq:kappax}\tilde \kappa =\sqrt{1+2x}-1\,. \end{equation}
We can reproduce the universal dimer limit $\kappa \to a_{13}^{-1}$ in the limit $a_{13} \gg r_0$ ($x\ll1$). 

It is interesting to note that in the above regime, the dimer energy is related to the difference in the well depths. Using Eqs.~\eqref{eq:finetuning} and \eqref{eq:kappax}, we have
\begin{equation} \label{eq:DeltaEta2} \Delta \eta^2 = \eta_{13}^2 - \eta_{12}^2 \approx 2 \tilde\kappa + \tilde\kappa^2. \end{equation}

Figure~\ref{fig:DimerEnergy} compares the dimer binding energy derived from the square well model with those by other theoretical methods. 
For the model parameters chosen above, one finds $E/h\approx -4.13$\,MHz, which is about 3\% deeper than the coupled channel calculation result. 

\begin{figure*}[t!]
\centering
\includegraphics[width=2\columnwidth]{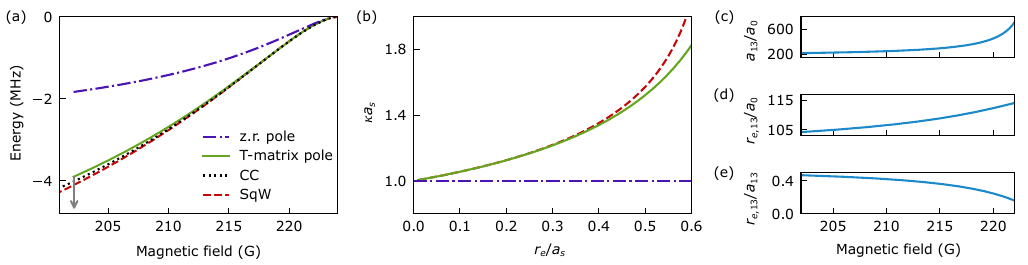}
\caption{ 
Effective-range effects on the $s$-wave dimer pole. 
(a) The dimer pole energy in the $13$ channel of $^{40}$K, versus field. The location of the $12$ scattering resonance, $B_0$, is indicated by the gray arrow on the axis. The pole energies approach zero at the $13$ resonance, roughly 22\,G higher than $B_0$.  
(b) Finite-range effect on the dimer pole: $\kappa$ versus $r_{e}/a_S$, for both the SqW theory (red dashed line) and the T-matrix pole (green solid line). 
(c) $a_{13}/a_0$ versus magnetic field. 
(d) $r_{e,13}/a_0$ versus magnetic field. 
(e) $r_{e,13}/a_{13}$ versus magnetic field. 
\label{fig:DimerEnergy}}
\end{figure*}

The dimer feature strength is given by the overlap between the $k_i=0$ scattering wave-function and that of the final dimer state, i.e., 
\begin{equation} I_d = |\bra{\kappa}\ket{k_i=0}|^2 = \ell_d\frac{C}{2 \pi}\,, \end{equation}
with the length scale
\begin{equation} \label{eq:SqWelld} 
\boxed{ \ell_d = \frac{1}{\kappa}\frac{(1+\tilde\kappa/2)^2}{1 + \tilde\kappa}}\,\,, \end{equation}
The correction to the $\Delta_d^\mathrm{zr}$ due to finite-range effects is then 
\begin{equation} \label{eq:AdSqW} 
A_d = \frac{\Delta_d}{\Delta_d^\mathrm{zr}} = \ell_d \kappa^2 a_{13} = \tilde\kappa \tilde a_{13} \frac{(1+\tilde\kappa/2)^2}{1 + \tilde\kappa} \end{equation}
We can compare the expression of $I_d$ to the shallow-dimer limit $\kappa r_0 \ll 1$, for which $I_d \to C/2 \pi \kappa$. For our parameters, $\kappa \ell_d\approx 1.09$.

Ref.~\cite{Baym:2007ki} showed that the total clock shift is given by $\hbar \Delta = n_2^{-1} \int d^3 r\,[v_{13}(\br) - v_{12}(\br)] \langle \psi_1^\dag(\br) \psi_2^\dag(0) \psi_2(0) \psi_1(\br) \rangle$, if one models the potential between two atoms in $\ket{i}$ and $\ket{j}$ by $v_{ij}(\br)$. Here $n_2$ is the number-density of $\ket{2}$ atoms, and $\langle \ldots \rangle$ measures the density-density correlator at $1-2$ separation $\br$. 
For our spherical square well model, $v_\sigma(\br)$ is only nonzero for $r<r_0$, and $\langle \ldots \rangle = |\phi_{12}(\br)|^2 / \mathcal{V}$, where the volume factor, $\mathcal{V}^{-1}$, comes from the center-of-mass degree of freedom and is canceled when dividing by $n_2 = N_2/\mathcal{V}$. Thus we have
\begin{equation} \hbar \Delta = N_2^{-1} (-V_{13} + V_{12}) \int_0^{r_0} dr |u_{12}(r)|^2\,, \end{equation}
where $u_{12} = \sqrt{4\pi} r \phi_{12,<}$. 
This is a general relation for the square well and demonstrates that the clock shift is convergent for any choice of parameters. If the depths $V_\sigma$ are different by a large energy, then the total clock shift is also large (but still finite).

The integral of the inner wave function can be found to be $2 \pi r_0 \mathcal{N}^2$ in the unitary case, and related to the contact parameter through Eq.~\eqref{eq:CSqWU}: 
\begin{equation} \hbar \Delta = \frac{r_0}{8 \pi N_2} (-V_{13} + V_{12}) C\,. \end{equation}
A low-energy clock shift can be guaranteed for the ``fine tuned'' scenario discussed above, in which $V_{13}$ is only slightly larger than $V_{12}$, sufficient to achieve the desired final-state scattering length $a_{13}$. This result can be written with reference to the zero-range \ $\Delta^\mathrm{zr}$, as 
\begin{equation} \label{eq:sqwuclockshift} 
\Delta = A_c \Delta^\mathrm{zr} \quad \mbox{,} \quad \Delta^\mathrm{zr} \equiv \frac{-\hbar} {2 \pi m a_{13}} \frac{C}{N}\,,  \end{equation} 
with 
\begin{equation} \label{eq:SqWAc}
\boxed{A_c = \tilde a_{13} \frac{\Delta \eta^2}{2}}\,.\end{equation}
Using Eq.~\eqref{eq:DeltaEta2}, this gives $A_c \approx 2.08$ for our parameters. 

At $1/a_{12}=0$, the rf spectrum for positive detuning $\omega>0$ is given by 
\begin{equation}
\Gamma(\omega)=\frac\pi2 \Omega_{23}^2\frac L\pi \int_0^\infty \! dk_f |\langle k_f|k_i=0\rangle|^2\delta(\omega-k_f^2/m)\,,
\end{equation}
with
\begin{equation}
|\langle k_f|k_i=0\rangle|^2=\frac{r_0^2 C }{2\pi L} \left[\frac{\Delta\eta^2 \cos(\tilde k_f+\delta_f)}{(\Delta\eta^2+\tilde k_f^2)\tilde k_f}\right]^2.
\end{equation}
Using Eq.~\eqref{eq:SqWBC}, one finds 
\begin{equation} \label{eq:SqWGamma} 
\Gamma(\omega) = \frac{\Omega_{23}^2 C \omega^{-3/2} }{8 \pi m^{1/2}} 
\big(1+ \frac{\omega \tan^2\!s}{s^2/mr_0^2}\big)^{-1} \frac{(\Delta \eta^2/mr_0^2)^2}{(\Delta \eta^2/mr_0^2 + \omega)^2} \,,
\end{equation}
with $s= \sqrt{\eta_{13}^2 + m r_0^2 \omega}$. 
The first two factors can be identified as Eq.~(1) from the main text, in the zero-range, deep-well limit. 
This can be seen from 
$\tan(s)/s \approx \tan(\eta)/\eta$ for $V_{13}^2 \gg \omega$, and $\tan(\eta)/\eta=1-\tilde a_{13}$ from Eq.~\eqref{eq:SqWaS}. For $a_{13} \gg r_0$, the zero-range case, then $(1-\tilde a_{13})^2 \approx \tilde a_{13}^2$. 
This gives us an clock-shift integral of the form 
\begin{equation} \begin{aligned}
\Delta_\mathrm{HFT}^\mathrm{zr} &= \frac{\int d\omega \Gamma \omega}{\int d\omega \Gamma} 
= \frac{1}{2 \pi^2 m^{1/2}} \frac{C}{N} \int \! d\omega \, \omega^{-1/2} \frac{1}{1+ \omega/\omega_a} \\
&= \frac{\sqrt{\omega_a}}{2 \pi m^{1/2}} \frac{C}{N} \,,
\end{aligned}\end{equation}
In the finite-range case, convergence of the integral relies on the third factor in Eq.~\eqref{eq:SqWGamma}. In the ``fine-tuned'' case, $\Delta \eta^2 = 2\tilde\kappa + \tilde\kappa^2$, so this cutoff is approximately at $2.2 \hbar^2/m r_0^2 \approx 15\,$MHz for our parameters. 
The integral can be resolved analytically by noting that in the unitary case $\eta_{12}=\eta_j$, for $\eta_{12}^2 \gg \Delta \eta^2$,
$\tan{s}/s \approx 2/(\Delta\eta^2 + m r_0^2 \omega)$ 
in the frequency regime $\omega \ll \eta_{12}^2/m r_0^2$. Even though this condition is eventually broken when integrating to infinite frequency, $\Gamma$ at frequencies far beyond $\Delta \eta^2/m r_0^2$ do not contribute significantly to the integral: 99\% of its value is obtained by $10^2/m r_0^2$, whereas $\eta_{12}^2 \sim 10^5$ for the parameter set given above. One then finds
\begin{equation} \begin{aligned} \label{eq:SqWAHFT}
A_\mathrm{HFT} &= \frac{\Delta_\mathrm{HFT}}{\Delta_\mathrm{HFT}^\mathrm{zr}} \\
&= \frac{\tilde a_{13}}{\pi} \int_0^\infty \! dy\, y^{-1/2} 
\frac{ (\Delta \eta^2 )^2 }{4y + (\Delta \eta^2 + y^2)^2 } \\
& = \boxed{ \frac{ \tilde a_{13} \, \Delta \eta^2}{2 \sqrt{1+ \Delta \eta^2}} }\,. \\
\end{aligned} \end{equation} 
For our parameters, $A_\mathrm{HFT} \approx 1.17$. 

If the HFT and dimer feature are the only aspects of the spectrum that determine the clock shift, then $\Delta = \Delta_d + \Delta_\mathrm{HFT}$. In relation to the zero-range shift, $\Delta = A_c \Delta^\mathrm{zr}$, $\Delta_d = 2 A_d\Delta^\mathrm{zr}$, and $\Delta_\mathrm{HFT} = A_\mathrm{HFT} \Delta^\mathrm{zr}$. Thus
\begin{equation} A_c \stackrel{!}{=} 2 A_d - A_\mathrm{HFT}\,. \end{equation}
One can verify that Eqs.~\eqref{eq:AdSqW}, \eqref{eq:sqwuclockshift}, and \eqref{eq:SqWAHFT} do satisfy this completeness condition, so long as Eq.~\eqref{eq:DeltaEta2} holds. 

\section{Sum rule \label{sm:Tmatrix}} 

The two-body scattering T-matrix is found by summing diagrams to all orders yielding (with $\hbar=1$)
\begin{equation} T(z) = \bar g \left( 1 - \frac{\bar g}{\mathcal{V}} \sum_p \frac{1}{z - p^2/m} \right)^{-1}\,, \end{equation}
where $\bar g$ is the bare (unrenormalized) coupling constant, $z = k^2/m + 0^+$ is the energy, $p$ and $k$ are collision momenta, $m$ is the atomic mass, and $\mathcal{V}$ is the volume. Using a momentum cutoff-off $\Lambda$ for the sum, one finds a renormalized coupling constant $g$, which we set $=4 \pi a/m$ where $a$ is the measured $s$-wave scattering length. $\Lambda$ is chosen such that the low-$k$ expansion of the T-matrix has the form 
\begin{equation} T^{-1} \approx \frac{m}{4 \pi} \left(-k \cot \delta + i k \right)\,, \end{equation}
with $\cot \delta \approx -a^{-1} + \frac12 r_e k^2$, where $r_e$ is the effective range. 
One finds that $\Lambda = 4 r_e^{-1}/\pi$, such that the T-matrix is 
\begin{equation} T^{-1}(k,a,r_e) = \frac{m}{4 \pi} \left( \frac{1}{a} + i k 
+ \frac{k}{\pi} \ln \frac{1 - \pi4 k r_e/4}{1 + \pi k r_e/4} \right)\,. \end{equation}


To find the shallow dimer in the final-state channel, we need to find the pole in its scattering T-matrix. Here, we take $k \to i \kappa$, with $\kappa >0$. 
One finds a nonlinear equation 
\begin{equation} \label{eq:pole}
\frac{1}{a_{13} \kappa} = 1 - \frac{2}{\pi} \tan^{-1}\left(\frac{\pi}{4} \kappa r_{e,13} \right)\,, \end{equation} 
whose small-$r_e$ limit is $\kappa \approx a_{13}^{-1}(1 + r_{e,13}/a_{13})$. 
Figure~\ref{fig:DimerEnergy} compares this pole to other approximations.  


We can use the clock-shift sum rule \cite{Punk2007,Baym:2007ki,Zhang:2008dc} and the same choice of cutoff to give a range-corrected clock shift. 
For a balanced gas, the total clock shift $\Delta$ is
\begin{equation} \label{eq:sumrule} 
\widetilde\Delta = \frac{\Delta}{E_F} = \frac{4}{m k_F} 
\frac{\bar g_{13}}{\bar g_{12}} \left(\frac{1}{g_{12}} - \frac{1}{g_{13}} \right) \widetilde C\,, \end{equation}
where $\widetilde C$ is defined as in the main text, and 
$g^{-1} = \bar g^{-1} + m \Lambda / 2 \pi^2$ for both channels. However, there is an ambiguity for which $r_{e,\sigma}$ to choose in determining the cutoff. 
For the unitarity case $a_{12}^{-1}=0$, one finds
\begin{equation} \label{eq:unitarity1} 
\widetilde\Delta
= -\left(1 - \frac{\pi^2}{8} \frac{r_{e,\sigma}}{a_{13}} \right)^{-1} \frac{\widetilde C}{\pi k_F a_{13}} \,. \end{equation}
In the zero-range limit, $|a_{13}| \gg r_{e,\sigma}$, we recover 
$\widetilde\Delta^\mathrm{zr}= -(\pi k_F a_{13})^{-1}{\widetilde C}$. The complete clock shift is then $\Delta = A_c \Delta^\mathrm{zr}$, where 
\begin{equation} \label{eq:AcTmatrix}
\boxed{ A_c = \left(1 - \frac{\pi^2}{8} \frac{r_{e,\sigma}}{a_{13}} \right)^{-1}}\,. \end{equation}
For our experimental scenario, one finds $A_c \approx 2.79$ for $r_{e,12}$ and $A_c \approx 2.36$ for $r_{e,13}$. Assuming that this range indicates a systematic uncertainty, we use $A_c = 2.58(22)$ in the main text. 

\section{Details of the experimental protocol \label{sm:experiment}}

{\em Preparation, thermometry and imaging.}
We start all experiments by preparing a spin-balanced mixture in a crossed optical dipole trap, providing harmonic trapping frequencies of approximately $\{170, 450, 500 \}$\,Hz in the $\{x, y, z\}$ directions. A typical gas contains about $N = 5(1) \times 10^4$ atoms of $^{40}$K with $E_F \approx 18\,$kHz at $T \approx 0.6T_F$. With further evaporation, our coldest measurements reach $T \approx 0.2T_F$ with $N \approx 1.3\times 10^4$ atoms and Fermi energy $E_F \approx 13$\,kHz. We can heat the gas to higher temperatures (up to $1.0 T_F$) by extinguishing the trap, allowing the atoms to expand for a variable interval (up to $0.8$\,ms), recapturing, and allowing thermalization before the measurement. These two techniques allow us to prepare gases in the range of temperatures shown in Fig.~\ref{fig:spectralweight}.

Thermometry is implemented with time-of-flight absorption imaging of the $\ket{1}$ state at 202.14\,G. The anisotropic cloud is interpreted assuming viscous expansion \cite{schaefer2010dissipative,cao2011universal} and an experimentally determined equation of state \cite{Ku_2012,VanHoucke2012}. We average many cycles of thermometry imaging to estimate the mean $N$, $E_F$ and $T/T_F$ for gases prepared for each measurement in Fig.~\ref{fig:spectralweight} to within 2\% statistical error.

State-selective time-of-flight imaging is performed at 209\,G, the zero-crossing of the 202.14\,G 12-Feshbach resonance, chosen to minimize interaction effects during time-of-flight. 
After spectroscopy at 202.14\,G, the magnetic field is jumped to 209\,G in $\sim 20\,\mu$s, via the discharge of a high-voltage capacitor into a secondary coil. Current in the primary coil is subsequently increased in 10\,ms to stabilize the total magnetic field at the same value. The state-selective imaging protocol is optimized for atoms in states $\ket{1}$ and $\ket{3}$, which necessitates exchanging the initial spin populations depending on measurement protocol; for non-dimer measurements, we exchange $\ket{1}$ and $\ket{2}$, whereas for the dimer measurement, we exchange $\ket{2}$ and $\ket{3}$. The exchange is performed via field-ramped adiabatic rapid passage in 9\,ms before again returning to the imaging field in 5\,ms. Including settling times throughout the procedure, the total process has a duration of ~26\,ms. 

{\em Linear response calibration.}
In the linear response regime, the transferred fraction $\alpha$ should be proportional to $\Omega_{23}^2$. We compensate for deviations by calibrating the saturation and correcting $\alpha$ to an extrapolated linear-in-$\Omega_{23}^2$ regime. This calibration is data-intensive, as it depends both on $\omega$ and on $T/T_F$: see Fig.~\ref{fig:saturation}. For each configuration, response is fit to $\alpha(\Omega_{23}^2) = \alpha_0(1-e^{-\Omega_{23}^2/\Omega_{0}^2})$, with free parameters $\Omega_{0}^2$ and $\alpha_0$. Figures in the main text report the response corrected for nonlinearities: $\alpha_0 \Omega_{23}^2/\Omega_{0}^2$. 

\begin{figure}[tb!]
\centering
\includegraphics[width=\columnwidth]{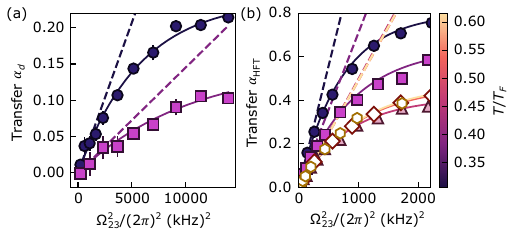}
\caption{
{\em Linear response calibration.} 
(a) Dimer transfer plotted against rf power. (b) HFT transfer for various rf power detuned $100$\,kHz from the 2-to-3 resonance (the detuning used in the HFT in Fig.~\ref{fig:spectralweight}). Solid lines are fits to the saturating transfer equation, and the dashed and dotted lines show extrapolated linear response. The color bar indicates the $T/T_F$ of the initial gas for each data set. 
\label{fig:saturation}}
\end{figure}

For contact measurements, we target $\alpha \lesssim 0.1$ for dimer and HFT regimes, and $\alpha \lesssim 0.03$ near resonance, to obtain adequate signal and yet remain near the linear response regime. The typical corrections to $\Gamma$ applied in Fig.~\ref{fig:spectralweight} of the main text were $\sim 1.3$ for the dimer and $\sim 1.1$ for HFT. 

{\em Inelastic alkali-dialkali collisions.}
After rf association, $\ket{13}$ dimers undergo an inelastic collision process $\ket{13}_\nu + \ket{2} \to \ket{13}_{\nu'} + \ket{2} + \mathrm{KE}$ \cite{quemener2012ultracold} where the dimer enters a lower vibrational state $\nu'$ by giving some kinetic energy $\mathrm{KE}$ to the reaction products, which are then lost from the trap. For an inelastic loss rate $\sim 10^{-10} \mathrm{cm}^3/\mathrm{s}$ \cite{quemener2012ultracold} and density $n = \sqrt{2} (mE_F)^{3/2} / (3\hbar^3 \pi^2) \sim 10^{13} \mathrm{cm}^{-3}$ for our typical $E_F$, this process occurs over a timescale $\tau \sim 1$\,ms. In an experiment cycle, clouds are held for $\sim 26$\,ms in the optical trap to prepare the cloud for imaging after rf spectroscopy. This is sufficient time for all dimers to undergo this vibrational relaxation process. The net effect is the loss of two $\ket{2}$ particles and one $\ket{1}$ particle for every dimer formed. This two-to-one loss is immediately apparent in our data, and insensitive to magnetic field (i.e., apparently unrelated to the $12$ Feshbach resonance). 

We measure the relaxation rate by ``swapping" the populations of unbound atoms in $\ket{1}$ and $\ket{2}$ at time delay $t_\pi$ after rf association of dimers. The final populations in $\ket{1}$ and $\ket{2}$ are then imaged. We find that $\delta N_2/\delta N_1$ decreases with $t_\pi$, which can be understood as follows. 
If the population swap is applied immediately after the dimer association pulse, the system is left as if we did not apply any pulse at all, and we observe greater loss in $\ket{2}$ compared to $\ket{1}$. Conversely, in the case of long $t_\pi$, the (initial) $\ket{2}$ atoms experience severe atom-dimer decay before being swapped into $\ket{1}$ for imaging. Fitting these data to an exponential, we find a best-fit decay time $\tau = 1.4(3)$\,ms for $T = 0.58(2) T_F$, and $\tau = 0.6(2)$\,ms for $T = 0.32(1) T_F$. We attribute this increase to the higher density of a colder gas. Both relaxation times are consistent with a vibrational decay cross-section comparable to the ``universal'' rate of dialkali collisional relaxation, described in Ref.~\cite{quemener2012ultracold} and references therein. We note that alkali relaxation dynamics are atypical, since ultracold atom-molecule relaxation rates are normally sensitive to the details of three-body interaction potentials.

{\em Dimer transfer signal in the spin channel.}
The resulting two-to-one loss after dimer association described in the previous section has a useful consequence. 
Measurements that rely upon the loss of atoms as a signal suffer from shot-to-shot number variation, since one cannot measure both the initial and final number of atoms in a single experimental cycle. 
Measuring instead the ratio of final number of atoms $R_{12} \equiv N_1/N_2$ reduces the effect of common-mode fluctuations. To obtain the fraction of atom pairs associated into dimers $\alpha_d$ from the number of atoms ratio, we assume the two-to-one loss ratio, such that $N_1 = N_1^\mathrm{ref} - \alpha_d N^\mathrm{ref}_\mathrm{tot}/2$ and $N_2 = N_2^\mathrm{ref} - \alpha_d N^\mathrm{ref}_\mathrm{tot}$ where $N^\mathrm{ref}_\mathrm{tot}$ is the initial total number of atoms.
Then, the ratio can be written $R_{12} = (N_1^{\mathrm{ref}}- \alpha_d N^\mathrm{ref}_\mathrm{tot}/2) / (N_2^{\mathrm{ref}} - \alpha_d N^\mathrm{ref}_\mathrm{tot})$. By solving for $\alpha_d$ and defining the fraction of atoms initially in state $\sigma \in {1,2}$ as $f_\sigma^\mathrm{ref} = N_\sigma^\mathrm{ref}/N^\mathrm{ref}_\mathrm{tot}$, we obtain
\begin{equation} \alpha_d = \frac{R_{12}f_2^\mathrm{ref} - f_1^\mathrm{ref}}{2R_{12} - 1}\,.\end{equation}
For a balanced cloud, the reference state fractions $f_\sigma^\mathrm{ref}$ are both 0.5. Since $\alpha_d$ can thus be found from $R_{12}$, its shot-to-shot fluctuation is reduced. 

{\em Feshbach-enhanced dipolar loss.}
The broad $p$-wave Feshbach resonance centered at 215\,G enhances dipolar relaxation in collisions between atoms in $\ket{1}$ and $\ket{3}$. Since $N_3 \ll \{ N_2, N_1 \}$ in our perturbative measurements, the effect of the collisional loss is only significant for atoms in $\ket{3}$. 

We correct for this by measuring the decay time $\tau_3$ at 209\,G (the field used for imaging), also finding that its dependence on temperature and number is significant. A typical $\tau_3$ is $100\,$ms, which we can understand as follows. Coupled channels calculations give an inelastic loss rate of $K_2 \approx 2 \times 10^{-12} \mathrm{cm}^3/\mathrm{s}$. The peak density in a harmonic trap is $n_{1, pk} = \sqrt{2} (mE_F)^{3/2} / (3\hbar^3 \pi^2) \sim 10^{13} \mathrm{cm}^{-3}$, giving $(K_2 n_{1, pk})^{-1}$ that is $\sim 30\,$ms, which is a time scale comparable to what is observed.

We calibrate lifetime curves and apply a corrective factor on $N_3$ of up to 30\% depending on the temperature of the initial system. This correction applies only to resonant and HFT transfer protocols. 

{\em Normalization of the spectrum.}
In the linear-response regime, the transfer rate of atoms obeys the sum rule 
\begin{equation}
\int \! d\omega \, \Gamma(\omega) = \Omega_{23}^2 \pi N/4 \, ,
\end{equation}
where, as in the main text, $\hbar\Omega_{23}/2$ is the matrix element in the rotating wave approximation. We can partition our data into the three phenomenological features, whose dimensionless spectral weights are 
\begin{equation} 
I_j \equiv \frac{ \int_j d \omega \, \Gamma }{ \int d \omega \, \Gamma} \,, \end{equation} 
normalized such that $\sum_j I_j = 1$. Here $j$ refers to the response and frequency range associated with the dimer feature ($I_d$), the resonant regime ($I_\mathrm{res}$), or the HFT ($I_\mathrm{HFT}$). 

The data shown in Fig.~\ref{fig:spectrum}(a) are taken at various temperatures and with pulse durations between $200\,\mu$s and 2\,ms. Direct integration shows that it does not satisfy the sum rule: $I_\mathrm{res} + I_\mathrm{HFT} + I_d \approx 0.52(2)$, which is inconsistent with unity. We note that the resonant $I_\mathrm{res} \sim 0.42$ provides most of the spectral weight, compared to $I_\mathrm{HFT} \sim 0.08(1)$ and $I_d \sim 0.02$ (here $T/T_F \approx 0.6$). 
In the following, we demonstrate that $I_\mathrm{res}$ is under-estimated, by comparing to an alternate protocol. 

In theory, the dynamical response to weak-drive coupling in a quantum many-body system consists of smoothly connected but physically distinct regimes: transition probability scales as $t_\mathrm{rf}^2$ at early times, and an FGR $t_\mathrm{rf}$-linear regime emerges at later time \cite{Haussmann:2009,Pieri:2009jp,Braaten2010,Randeria2010}. Our analysis to find $\Gamma$ assumes a FGR response. In order to spectrally resolve the resonant peak ($\sim 10\,$kHz) and remain within the FGR regime \cite{Navon:2025b}, the resonant peak was probed with 2\,ms pulses. However, this pulse length exceeds the spin-coherence of our system, which is limited by magnetic-field stability to $\sim 500\,\mu$s. While shorter pulse times can be (and are) used for the HFT, we could not use a shorter pulse time near resonance without leaving the FGR regime. In other words, the magnetic-field noise does not allow the central peak to be well resolved. This prevents an accurate determination of the sum rule. 

\begin{figure}
    \centering
    \includegraphics[width=\linewidth]{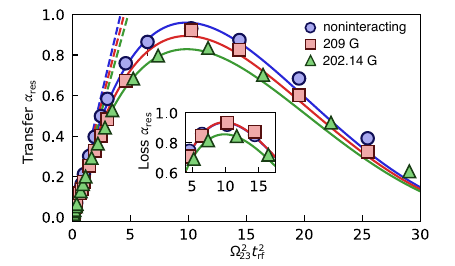}
    \caption{
    {\em Short-pulse spectral weight.} 
    Resonant transfer $\alpha_\mathrm{res}$ for various initial systems probed by a square pulse of duration $t_\mathrm{rf} = 10\,\mu$s in the Rabi oscillation regime. Data is taken for various initial systems, solid lines are fits to Eq.~\ref{eq:Zfactor}, and dashed lines are leading-order extrapolations. Imaging a noninteracting, spin-polarized system acts as reference for imaging systematics that reduce fidelity; here $\zeta = 0.96(1)$ (blue circles). $\zeta$ is further reduced for a noninteracting balanced spin mixture measured at $209\,$G (red squares) due to the influence of $p$-wave loss lowering the apparent count of $N_3$. Green triangular data is taken for a unitary balanced spin mixture at $202.14\,$G, where $\zeta$ decreases again because of reduction of the spectral weight on resonance. Comparing red and green estimates the resonant spectral weight $I_\mathrm{res} = 0.93(2)$.
    {\bf{Inset:}} Analysis of loss in $N_2$ shows agreement between blue and red data due to lack of $p$-wave loss whereas reduction in $\zeta$ caused by interactions (green) is still apparent.
    }
    \label{fig:SM_Zfactors}
\end{figure}

An alternate measure of $I_\mathrm{res}$ can be obtained in the short-pulse, $t_\mathrm{rf}^2$-transfer regime. If one considers a resonant pulse where $1/t_\mathrm{rf}$ dominates over the resonant spectral width (and also neglects the weight of the convolved HFT portion, whose peak value is already about $\sim 10^2$ times smaller than the peak value on resonance due to $\omega^{-3/2}$ decay), the transfer fraction will go as
\begin{equation} \label{eq:Zfactor}
\alpha_\mathrm{res} = \zeta \sin^2\left(\frac{\Omega_{23}t_\mathrm{rf}}{2}\right)\,,
\end{equation}
where $\zeta$ is the weight of spectral region probed by the pulse. $\zeta=1$ for a noninteracting gas (and with properly calibrated imaging). 
A reduced $\zeta < 1$ is expected for an interacting system because spectral weight transfers from the near-resonant feature to the HFT and dimer features. The depletion of the single-particle feature increases with correlation strength, since $I_d$ and $I_\mathrm{HFT}$ increase linearly with $C$. 

We extract $\zeta$ through a fit to $\alpha_\mathrm{res}$ measured with a fixed $t_\mathrm{rf}$ satisfying the short-pulse condition and a variable $\Omega_{23}^2$. Transfer data to $\ket{3}$ taken for various initial systems are shown in Fig.~\ref{fig:SM_Zfactors}, where the solid line fits are to Eq.~\ref{eq:Zfactor} and the dashed lines are leading-order extrapolations to $\alpha_\mathrm{res} \to A \Omega_{23}^2 t_\mathrm{rf}^2 /4$. The blue circles depict transfer for an initially spin-polarized and noninteracting $\ket{2}$ cloud and we find $\zeta^\mathrm{ideal} = 0.96(1)$ demonstrating agreement with theory within a $4\%$ imaging systematic. Data for a noninteracting spin mixture of $\ket{1}$ and $\ket{2}$ are represented by the red squares; here $\zeta^\mathrm{n.i.}=0.89(1)$ due to under-counting of $N_3$ caused by $p$-wave loss at 209\,G mentioned earlier. In the inset, the same data is displayed when analyzed for loss in $N_2$ as also described in the main text; $1 - N_2/\bar{N}_2^{\mathrm{ref}}$ -- because state $\ket{2}$ is negligibly affected by $p$-wave loss, the amplitudes between blue and red datasets match with excellent agreement. Lastly, the green triangles are taken for a balanced mixture at unitary, which is affected both by $p$-wave loss and a reduction in $\zeta$ caused by interactions resulting in $\zeta^\mathrm{int}=0.83(1)$. Assuming the $p$-wave loss rate is similar to the noninteracting case, we can estimate the resonant spectral weight $I_\mathrm{res}$ through the ratio $\zeta^\mathrm{int}/\zeta^\mathrm{n.i.}$, which is $0.93(2)$. Combined with the estimated $I_\mathrm{HFT}$ and $I_d$ from earlier in this discussion, the total spectral weight becomes reasonably close to 1. The exact ratio $\zeta^\mathrm{int}/\zeta^\mathrm{n.i.}$ depends on the temperature and density of the gas, however at this level we conclude that (a) The spectral weight extracted through a resonant short pulse is close to unity as expected with some reduction due to interactions; (b) $I_\mathrm{res}$ remains the dominant contribution to $I$; (c) The remaining spectral weight most likely goes to the HFT and dimer features; and (d) Though we cannot exclude the possibility of there being other spectral features missing from the analysis, we expect their contributions to be small.

\section{Case of Lithium} 

This section considers rf spectroscopy $^6$Li in a spin mixture of the lowest $\ket{a}$ and third-lowest $\ket{c}$ ground states, at the 689.7\,G Feshbach resonance for $s$-wave interactions between them. For spectroscopy on the $\ket{a}$-to-$\ket{b}$ transition, the final-state scattering length (between $\ket{b}$ and $\ket{c}$) is $a_\mathrm{f} \approx 1170\,a_0$, with effective ranges $r_\mathrm{e,i} \approx 86.1\,a_0$ and $r_\mathrm{e,f} \approx 78.0\,a_0$. Here ``i'' refers to interactions between $\ket{a}$ and $\ket{c}$, and ``f'' refers to interactions between $\ket{b}$ and $\ket{c}$. (We have changed notation in this section to avoid confusion, since the energy-ordering of $^6$Li states differs from the $^{40}$K case discussed in the main text.) The value for $a_\mathrm{f}$ is taken from the parameterization in Ref.~\cite{Jochim:2013}, while the values of $r_\mathrm{e,i}$ and $r_\mathrm{e,f}$ are from a new coupled-channels calculation. 

Since $a_\mathrm{f}>0$, a dimer resonance appears in the rf spectrum. Its binding energy is anticipated by Eq.~\eqref{eq:pole} to be $\omega_d \approx 2 \pi \times 0.472$\,MHz, agreeing with the coupled-channels pole to within 1\,kHz. This is comparable to the observed transfer peak reported in Ref.~\cite{Navon:2023}, at $0.464(1)$\,MHz, which agrees with earlier work~\cite{Ketterle:2008}. The final-state binding energy is closer to universal ($\kappa a_\mathrm{f} \approx 1.03$) than the case considered for $^{40}$K in the main text. The universal expectation for the dimer-strength contact relation would be $\ell_d = \kappa^{-1} \approx 1130\,a_0$. 

The multichannel matrix element of the dimer feature is dominated by the $ac$-to-$bc$ term, with a sub-leading $bd$-to-$bc$ term that interferes constructively, boosting dimer weight by $\sim 10\%$. This yields a predicted dimer length $\ell_d = 1220(20)\,a_0$, and $\kappa \ell_d = 1.08(2)$. By comparison, the SqW-model modification of the dimer weight is insignificant: $\kappa \ell_d - 1 \approx 10^{-3}$ from ~\eqref{eq:SqWelld} (and also, $A_\mathrm{HFT} - 1 < 10^{-3}$ from Eq.~\eqref{eq:SqWAHFT}). 
This scenario echoes the case of $^{40}$K: the dimer weight is more sensitive to multichannel effects than one might expect from its binding energy. 
Note that the SqW model here is tuned to the $a_\mathrm{f}$ and $r_\mathrm{e,i}$ found from CC, but without another adjustable parameter, it gives an $r_\mathrm{e,f}$ that is $6\%$ larger than the CC value. This should be a higher-order correction since $r_\mathrm{e,f}/a_\mathrm{f} < 0.1$. 
The finite-range correction to the total clock shift is larger, but still close to unity: $A_c \approx 1.06$ from the SqW model, \eqref{eq:SqWAHFT}, and $A_c \approx 1.07$ from the T-matrix model, Eq.~\eqref{eq:AcTmatrix}. 

With this $\ell_d$, the dimer feature strength would be $I_d \approx 0.1 \widetilde{C}$ for $E_F/h = 15\,$kHz. 
As in the case of $^{40}$K, probing on the dimer feature should provide a rapid probe for contact measurement. For $t_\mathrm{rf} = \tau_F$, the power-efficiency figure of merit $\tau_F^2 \Omega^2/\alpha$ is $4 \pi (k_F \ell_d \tau_F^2 \widetilde C)^{-1} \sim 45 \widetilde{C}^{-1}$. By comparison, for HFT, $\tau_F^2 \Omega^2/\alpha$ is unchanged from the case of $^{40}$K, i.e., $8 \pi R^3 \widetilde{C}^{-2} \sim 10^6 \widetilde{C}^{-2}$. The improved efficiency may be even more helpful for $^6$Li, since in the Paschen-Bach regimes, the lowest three states have relatively small magnetic dipole matrix elements. 

Even with unlimited Rabi frequency, the structure of the HFT will limit the speed at which the contact can be probed. One must use $\omega < \omega_a$ to avoid the $\Gamma \propto \omega^{-5/2}$ regime, since this would scale more slowly than the single-particle $\omega^{-2}$ contamination. For the example numbers here, $\tilde \omega_a \approx 30$. Requiring that $\alpha_\mathrm{HFT}/\bar\alpha_\mathrm{res} \geq 10$, one finds that $t_\mathrm{rf} \gtrsim 8 \tau_F \widetilde C^{-1}$ is required. 

Probing on the dimer feature, pulse durations are restricted to $t_\mathrm{rf} \gtrsim \omega_d^{-1} (2 \pi R)^{1/2} (k_F \ell_d \widetilde C)^{-1/2}$ for the same purity constraint. This gives $t_\mathrm{rf} \gtrsim 5\,\mu$s, or $t_\mathrm{rf} \gtrsim 0.5 \tau_F \widetilde C^{-1/2}$. For ensembles $\widetilde C$ of order unity, using rf dimer projection can be an order of magnitude faster than probing the HFT. 

\end{appendices}
\fi
\end{document}